\documentclass{article}
\usepackage{amsmath,amssymb,amsfonts,textcomp}
\usepackage{graphicx}
\usepackage{color}
\usepackage{float}

\definecolor{red_ex}{rgb}{0.7,0,0}
\definecolor{blue_ex}{rgb}{0,0,0.7}
\newcommand{\bleu}[1]{\textbf{\textcolor{blue_ex}{#1}}}

\begin{document}

\title{Improving Update Summarization by \\ Revisiting the MMR Criterion}
       
%\author[udem]{Florian Boudin\corref{cor1}} 
%\ead{boudinfl@iro.umontreal.ca} 
%\author[lia]{Juan-Manuel Torres-Moreno} 
%\ead{juan-manuel.torres@univ-avignon.fr} 
%\author[lia]{Marc El-B\`eze} 
%\ead{marc.elbeze@univ-avignon.fr} 
%\address[lia]{Laboratoire Informatique d'Avignon, 339 chemin des Meinajaries, BP1228, 84911 Avignon Cedex 9, France}
%\address[udem]{DIRO, Université de Montréal, CP. 6128, succursale Centre-ville, Montréal (Québec), H3C 3J7, Canada}

\author{Florian Boudin$^1$, Juan-Manuel Torres-Moreno$^{1,3}$ and Marc El-B\'eze$^1$\\
$^1$ DIRO, Universit\'e de Montr\'eal, CP. 6128\\succursale Centre-ville, Montréal (Qu\'ebec), H3C 3J7, Canada\\
$^2$ Laboratoire Informatique d'Avignon\\
BP 1228 84911 Avignon Cedex 09, France\\
$^3$ Ecole Polytechnique de Montr\'eal, CP. 6128\\succursale Centre-ville, Montréal (Qu\'ebec), H3C 3J7, Canada
}

\maketitle

\begin{abstract}
This paper describes a method for multi-document update summarization that relies on a double maximization criterion.
A Maximal Marginal Relevance like criterion, modified and so called \textsc{Smmr}, is used to select sentences that are close to the topic and at the same time, distant from sentences used in already read documents.
Summaries are then generated by assembling the high ranked material and applying some ruled-based linguistic post-processing in order to obtain length reduction and maintain coherency.
Through a participation to the Text Analysis Conference (TAC) 2008 evaluation campaign, we have shown that our method achieves promising results.
\end{abstract}

% Section Introduction
\section{Introduction}
\label{sec:intro}

Text summarization is the process of automatically creating a compressed version of a given text that provides useful information for the user \cite{erkan2004}.
Query-oriented summaries focus on a user's need, and extract the information related to the specified topic given explicitly in the form of a query \cite{daume_thesis}.
On the other hand, generic summaries try to cover as much as possible the information content.
%Multi-document summarization is a task devised for creating a summary out of a collection of documents on a focused topic .
%In query-focused summarization, this topic is given explicitly in the form of a user's query.
Over the past few years, extensive experiments on query-oriented multi-document summarization have been carried out.
%Most of the strategies to produce summaries are based on an extraction method, which identifies salient textual segments, most often sentences, in documents.
Extractive summarization produces summaries by choosing a subset of sentences in the original documents.
Sentences are then ordered and assembled according to their relevance to generate the summary \cite{mani1999}.
This contrasts with abstractive summarization that involves rephrasing information in the text.
Although human beings typically produce summaries in an abstractive way, most of the research is on extractive summarization.
This is due to the fact that tools needed to construct semantic representations or generate natural language have not reached a mature stage today.
Moreover, existing abstractive summarizers often depend on an extractive component.
For example, \cite{radev1998} use a language generation component on top of a multi-document extractive summarizer to produce the final summary.
In this paper, we focus on query-oriented multi-document text summarization, where the goal is to produce a summary of multiple documents about a specified topic.

With the ever increasing popularity of news search engines, displaying the information in a more practical and pleasant way is becoming a challenging and important issue.
One possible solution is to summarize multiple news so as to propose only one short text instead of raw aggregated headlines.
This is, intuitively, a reasonable solution though producing summaries from large collection of documents is a very complicated task.
However, as the number of documents increases, facts that are considered as important --and have to appear in the summary-- also become more numerous.
In this case, a choice must then be made to drop important facts in order to satisfy size constraints.
One way to tackle this problem is to remove facts that the user is already aware of.
This variant of text summarization is called \emph{update summarization}.
More formally, update summarization is the task of producing summaries while minimizing redundancy with previously read documents (from now on \emph{history}).

Recently introduced at the Document Understanding Conference (DUC) 2007\footnote{Document Understanding Conferences are conducted since 2000 by the National Institute of Standards and Technology (NIST), http://www-nlpir.nist.gov}, update summarization is an emerging summarization task that brings new challenges to sentence ranking algorithms.
Indeed, segments have to be selected according to their salience but also to their ability to capture novelty.
Existing approaches are derived from state-of-the-art query-oriented multi-document summarizers by the addition of some constraints about redundancy and novelty detection.
These include Machine Reading \cite{hickl_duc07}, graph-based summarization \cite{li_coling08,witte_duc07}, Maximal Marginal Relevance (MMR) \cite{lin_duc07}, and novelty boosting \cite{boudin_ranlp07}.
The fact that most of them are relying on linguistic resources or tools such as taggers and parsers is a limiting factor for the adaptation to other languages or domains.

In this paper we propose a sentence ranking algorithm inspired by the well known MMR re-ordering algorithm.
Sentences are scored thanks to a double maximization criterion that strives to maximize sentence's relevance while maximizing non-redundancy with the previously read documents.
Our formulation combines word-level similarity measures in an information retrieval approach, ranking sentences by their similarity to the topic and the (inverse) similarity to other sentences in history.
We show that our method, although using minimal linguistic resources, can achieve good results among state-of-the-art summarizers. %and also performs consistently through different languages (English and French).
Preliminary results about the sentence re-ranking process were published in \cite{boudin_coling08,boudin_thesis}.
The remainder of this paper is organized as follows.
An overview of related work is provided in Section \ref{sec:relatedwork}.
Section \ref{sec:method} presents our three steps summarization method: pre-processing, sentence ranking and linguistic post-processing.
Experimental results are presented in Section \ref{sec:experiments}, followed by discussions and conclusions.

% Section Related Work
\section{Related Work}
\label{sec:relatedwork}

Introduced by Luhn in the fifties \cite{luhn1958}, research on automatic summarization can be qualified as a long tradition.
In the strategy proposed by Luhn, source sentences are scored for their component word values as determined by tf*idf-type weights.
Scored sentences are then ranked and selected from the top until some summary length threshold is reached.
Finally, the summary is generated by assembling the selected sentences in original source order.
Although fairly simple, this extractive methodology is still used in current approaches.
Later on, \cite{edmundson1969} extended this work by adding simple heuristic features of sentences such as their position in the text or some key phrases indicating the importance of the sentences.
As the range of possible features for source characterization widened, choosing appropriate features, feature weights and feature combinations have became a central issue.
A natural way to tackle this problem is to consider sentence extraction as a classification task.
To this end, several machine learning approaches that uses document-summary pairs have been proposed \cite{kupiec1995,teufel1997}.
Summarization then started gaining more momentum with the SUMMAC\footnote{TIPSTER Text Summarization Evaluation Conference  (SUMMAC) conducted in May 1998, http://www-nlpir.nist.gov/related\_projects/tipster\_summac/index.html} evaluation \cite{mani2002}, followed by the DUC evaluation conferences.

New tasks have been continuously added to the summarization issue as approaches became more robust and resources grew larger.
\cite{allan2001} were amongst the first to tackle the update summarization problem.
Their approach, originally developed as a tool to monitor changes in news coverage over time, uses topic detection and tracking techniques to determine which sentences capture usefulness and novelty.
The most intuitive way to go about update summarization would be to be identify temporal references within documents (dates, elapsed times, temporal expressions, etc.) and to construct a timeline of the events.
It is a complex task as temporal references depend on surrounding elements in the discourse but also require an understanding of the ontological and logical foundations of temporal reference construction \cite{Hinrichs1986}.
Assuming the timeline is constructed, update summaries could be produced by assembling sentences containing the most recent events.
However, most recently written material is not necessarily latest facts.
This way, focusing the summaries on information that the user is not aware of can be seen as identifying unseen facts.
Existing approaches rely exclusively on content-based redundancy removal without recourse to temporal detection.
\cite{hickl_duc07} propose a machine reading method to construct knowledge representations from clusters of documents.
Sentences that are containing “new” facts (i.e. that could not be inferred by any document from the history) are selected to generate the summary.
%Even though this approach achieves good results (best system at the duc 2007 update task), it requires very large linguistic resources.
A rule-based method using fuzzy coreference cluster graphs was introduced by \cite{witte_duc07}.
This approach can be applied to various summarization tasks but requires to manually write the sentence ranking scheme.
\cite{boudin_ranlp07} first use a na\"{\i}ve similarity ratio to select sentences that are relevant and dissimilar to sentences from history.
On top of this ranking approach, a second method called novelty boosting is used.
The latter extends the topic by the unique terms in the cluster, thus biasing the ranking towards maximizing relevance not only with respect to the topic, but also to the novel aspects of the topic in the cluster.

%However, several research use
%In contrast of these methods, most of the research relies on the informationnal content of sentences to remove redundancy.

% Section Method
\section{Method}
\label{sec:method}

In this section we present the details of the proposed text summarization method.
As mentioned earlier, our work models sentence ranking as a double maximization criterion.
We define $H$ to represent the previously read set of documents (history), $Q$ to represent the query and $s$ the candidate sentence.
The following subsections formally define document pre-processing, the sentence scoring method and the summary generation process.

\subsection{Pre-processing}

The first step is to prepare documents for the ranking process.
As we use extractive summarization, documents have to be chunked into cohesive textual segments that will be assembled to produce the summary.
The importance of pre-processing is predominant because the selection of segments is based on words they contains \cite{ledeneva_micai08}.
The choice was made to split documents into full sentences, in this way obtaining textual segments that are likely to be grammatically correct.
Afterwards, sentences are going through several basic normalization steps in order to reduce computational complexity.
An example of document pre-processing is given in Table \ref{tab:pretraitement}.
The process is composed by the following steps: \\

\begin{enumerate}

\item \textbf{Sentence splitting}: a simple rule-based method is used for sentence splitting\footnote{The software is available from http://duc.nist.gov/duc2004/software/.}. Documents are chunked at the dot, exclamation and question mark signs. Prior to that, ambiguous composed person names (i.e. ``George W. Bush") are detected to reduce segmentation errors. 
\newline

\item \textbf{Sentence filtering}: words are converted to lowercase and cleared up from sloppy punctuation. Words that do not carry meaning such as functional or very common words are removed.
\newline

\item \textbf{Date normalization}: dates are rewritten and extended with time related words. For example, ``december 15, 1982" is replaced by ``12/15/1982" and enriched with ``\_december\_ \_1982\_". Standardized dates allow to minimize the scoring function bias (i.e. considering only one word for one concept instead of three in this example) while enrichment is useful to link facts that were happening at the same period of time (month or year).
\newline

\item \textbf{Word normalization}: remaining words are replaced by their simplified forms (i.e. inflected forms ``go", ``goes", ``went", ``gone"... are replaced by ``go") using a word root database ($\approx$ 88 000 entries). In case of ambiguity, the most frequent word is chosen.

\end{enumerate}

\begin{table}[h!]
\centering
\begin{tabular}{|p{.008\textwidth}|l|}
\hline

\hspace*{-0.5em} \rotatebox{90}{\hspace*{-1.8em} Original} &

\begin{minipage}{0.93\textwidth}
\vspace*{0.2em}
\small{\textbf{WASHINGTON \_}$^{\textbf{(1)}}$ A federal judge Monday found President Clinton in civil contempt of court for lying in a deposition about the nature of his sexual relationship with former White House intern Monica S. Lewinsky. Clinton, in a January 1998 deposition in the Paula Jones sexual harassment case, swore that he did not have a sexual relationship with Lewinsky. Clinton later explained that he did not believe he had lied in the case because the type of sex he had with Lewinsky did not fall under the definition of sexual relations used in the case. \vspace*{-0.5em} \\ 
}
\end{minipage} \\

\hline

\hspace*{-0.5em} \rotatebox{90}{\hspace*{-2em} Splitted} &
\begin{minipage}{0.93\textwidth}
\vspace*{0.2em}
\small{\textcolor{blue}{$<$s0$>$}A federal judge Monday found President Clinton in civil contempt of court for lying in a deposition about the nature of his sexual relationship with former White House intern Monica S. Lewinsky.\textcolor{blue}{$<$/s0$>$}$^{\textbf{(2)}}$ \\
\textcolor{blue}{$<$s1$>$}Clinton, in a January 1998 deposition in the Paula Jones sexual harassment case, swore that he did not have a sexual relationship with Lewinsky.
\textcolor{blue}{$<$/s1$>$} \\
\textcolor{blue}{$<$s2$>$}Clinton later explained that he did not believe he had lied in the case because the type of sex he had with Lewinsky did not fall under the definition of sexual relations used in the case.\textcolor{blue}{$<$/s2$>$} \vspace*{-0.5em} \\ 
}
\end{minipage} \\
\hline

\hspace*{-0.5em} \rotatebox{90}{\hspace*{-2.2em} Processed} &

\begin{minipage}{0.93\textwidth}
\vspace*{0.2em}
\small{\textcolor{blue}{$<$p0$>$}federal judge monday find$^{\textbf{(3)}}$ president clinton civil contempt court lie deposition nature sex relation former white house intern monica lewinsky\textcolor{blue}{$<$/p0$>$}$^{\textbf{(4)}}$ \\
\textcolor{blue}{$<$p1$>$}clinton 01\_1998 \_january\_ \_1998\_$^{\textbf{(5)}}$ deposition paula jones sex harassment case swear sex relation lewinsky\textcolor{blue}{$<$/p1$>$}\\
\textcolor{blue}{$<$p2$>$}clinton late explain believe lie case type sex lewinsky fall define sex relation use case\textcolor{blue}{$<$/p2$>$}\vspace*{-0.5em} \\ 
}
\end{minipage} \\
\hline

\end{tabular}
\caption{Example of pre-processing applied to the document NYT19990412.0403 from cluster D0646A of DUC 2006. News agency name is removed (1); document is segmented into sentences (2); words are normalized (3); punctuation and case are removed (4); dates are standardized end enriched (5).}
\label{tab:pretraitement}
\end{table}

\subsection{Ranking}

Sentences are scored according to the fact that they contain material satisfying the need formulated in the user's query.
Ranking sentences for query-oriented summarization can be seen as a passage retrieval task in information retrieval.
In this paradigm, sentences sharing most of their vocabulary with the query are likely to be informational for the reader.
Each sentence is then scored by computing a combination of two similarity measures with the query.
The first similarity measure is the well known $cosine$ \cite{salton1975} computed on the sentence and the query vectorial representations in the document’s term-space (denoted respectively $\vec{s}$ and $\vec{Q}$).
The decision was made not to use the classical $tf \times idf$ weighting scheme \cite{jones1972} because of the difficulty to find similar data and generate pertinent weight lists. 
The main weakness of $cosine$ and more generally of all similarity measures using words for tokens is that they are relying too much on term normalization.
Their performance dramatically decreases with wrongly or non normalized words.
That is why we propose a second similarity measure based on the Jaro-Winkler distance \cite{winkler1999} that can bridge morphologically similar words in order to smooth normalization and misspelling errors.
This measure can be classified as an improved edit distance between two word sequences.
The Jaro-Winkler distance, denoted \textsc{Jw}, calculates the number of operations required to transform a string into another one.
It uses the number of matching characters and transpositions to compute a similarity score between two terms, giving more favourable ratings to terms that match from the beginning.
Originally introduced to tackle normalization issues in automatic summarization of chemistry articles \cite{boudin_gotal08}, this distance was extended to compute a similarity measure between a sentence $s$ and the query $Q$:

\begin{equation}
\textsc{Jw}_e(s,Q) = \frac{1}{|Q|} \cdot \sum_{q \in Q} \max_{m \in S^\prime}  \textsc{Jw}(q,m)
\label{eq:jaro}
\end{equation}
\vspace*{0.1em}

\noindent where $S^\prime$ is the term set of $s$ in which the terms $m$ that already have maximized $\textsc{Jw}(q,m)$ during the previous steps of the summation are removed.
The final score is calculated using a linear combination of the two similarity measures.
Equation \ref{eq:sim} shows how to compute the relevance score between a sentence $s$ and a query $Q$.

\begin{equation}
Sim_1(s, Q) = \alpha \cdot cosine(\vec{s}, \vec{Q}) + (1 - \alpha) \cdot \textsc{Jw}_e(s,Q)
\label{eq:sim}
\end{equation}
\vspace*{0.1em}

The Maximal Marginal Relevance (MMR) algorithm \cite{carbonell1998} has been successfully used in query-oriented summarization \cite{ye2005}.
It strives to reduce redundancy while maintaining query relevance in selected sentences.
The summary is constructed incrementally from a list of ranked sentences, at each iteration the sentence which maximizes MMR is chosen:

\begin{equation}
\textsc{MMR}  = \underset{s \in S}{\arg\max} \ [\ \lambda \cdot Sim_1(s,Q)  - ( 1 - \lambda)\ \cdot \max_{s_j \in E} Sim_2(s, s_j)\ ]
\label{eq:mmr}
\end{equation}
\vspace*{0.1em}

\noindent where $S$ is the set of candidates sentences and $E$ is the set of selected sentences.
$\lambda$ represents an interpolation coefficient between relevance and redundancy.
In the original formulation, $Sim_1$ and $Sim_2$ were computed using the $cosine$ similarity measure.
Although this measure has been proven to be efficient, any other similarity measure between sentences remains appropriate.

We propose an interpretation of MMR to tackle the update summarization issue.
Unlike previous work such as \cite{lin_duc07}, our approach does not require iterative re-ranking.
To remove sentences containing redundant material, the set of selected sentences $E$ is replaced by the set of sentences in history.
In terms of computational complexity, this means that each candidate is compared to all sentences from $H$.
Since $Sim_1$ and $Sim_2$ are ranged in $[0,1]$, they can be seen as probabilities even though they are not.
This way, $Sim_1$ is considered as the probability to be relevant to the topic and $Sim_2$ as the probability to be redundant with history.
We propose to rewrite (\ref{eq:mmr}) by adding the constant $(1 - \lambda)$ as (\textsc{NR} stands for Novelty Relevance):

\begin{align}
\textsc{NR}\ & =\  \underset{s \in S}{\arg\max} \ [\ \lambda \cdot Sim_1(s,Q) + (1 - \lambda) - ( 1 - \lambda) \ \cdot \max_{s_h \in H} Sim_2(s, s_h)\ ] \nonumber \\[5pt]  
                   & =\  \underset{s \in S}{\arg\max} \ [\ \lambda \cdot Sim_1(s,Q) + ( 1 - \lambda)\ \cdot (1 - \max_{s_h \in H} Sim_2(s, s_h))\ ] \label{eq:nr1}
\end{align}
\vspace*{0.1em}

This makes more sense because it combines relevance and non-redundance instead of focusing on redundancy penalization.
According to our intuition, we presume that (\ref{eq:nr1}) is more or less corresponding to an \textsc{or} ($\vee$) combination.
But we are obviously looking for a criterion corresponding to \textsc{and} ($\wedge$).
Since the similarities are independent, we can use the product combination.
%We can understand that (\ref{eq:mmr2}) equates to an \textsc{or} ($\vee$) combination.
%But as we are looking for a more intuitive \textsc{and} ($\wedge$) and since the similarities are independent, we have to use the product combination.
Sentences are scored thanks to a double maximization criterion in which the best ranked one will be the most relevant to the query \textsc{and} the most different to the sentences in $H$:

\begin{equation}
\textsc{Smmr}(s) = Sim_1(s,Q) \cdot \left (1 - \max_{s_h \in H} Sim_2(s, s_h)\right ) ^{{\mathcal{N}\hspace*{-0.08cm}f(H)}}
\label{eq:smmr}
\end{equation}
\vspace*{0.1em}

Decreasing parameter $\lambda$ in (\ref{eq:mmr}) with the length of the summary was suggested by \cite{murray2005} and successfully used in the DUC 2005 by \cite{hachey2005}, thereby emphasizing the relevance at the outset but increasingly prioritizing redundancy removal as the process continues.
Similarly, we propose to follow this assumption in \textsc{Smmr} using a function denoted $\mathcal{N}\hspace*{-0.08cm}f$ that as the amount of data in history increases, prioritizes non-redundancy ($\mathcal{N}\hspace*{-0.08cm}f(H) \rightarrow 0$).
We have defined this parameter function $\mathcal{N}\hspace*{-0.08cm}f$ as ``novelty factor". 

A special breed of redundancy is proliferating in news articles as journalists increasingly rely on the fact that news articles have to be as universally understandable as possible.
This means that most of the news articles contain previous facts and/or pointers to previous articles in order for a reader, that does not know anything on the subject, to catch on.
This is why we think that a normalized Longest Common Substring (LCS) measure between two sentences is well adapted to be used as the non-redundancy measure ($Sim_2$).
For example, LCS can easily detect sentence rewritings, specially when the sentence is structured around a redundant sub-sentence.

% Sous-section Post-Processing
\subsection{Post-processing}
\label{sec:post-processing}

Once sentences are selected to be assembled in the final summary, some linguistic treatments are applied.
Indeed, once out of their contexts, discursive forms are considerably decreasing summary's coherence.
For example, two sentences one next to the other in the summary may be in opposition while not dealing with the same subject.
Our rule based linguistic post-processing targeted sentence length reduction and coherency maximization.
An example of summary post-processing is given in Table \ref{tab:posttraitement}.
The process is composed by the following steps:
\newline

\begin{enumerate}
\item \textbf{Acronym rewriting}: first occurrence of an acronym is replaced by its complete form (acronym and definition); following ones only by their reduced forms. Definitions are automatically mined in the corpus by pattern matching. In case of acronym ambiguity, the most frequent one is selected.
\newline

\item \textbf{Date and number rewriting}: numbers are reformatted and dates are normalized to the US standard forms (\textsc{mm/dd/yyyy}, \textsc{mm/yyyy} and \textsc{mm/dd}).
\newline

\item \textbf{Temporal references rewriting}: time tags are used to replace fuzzy temporal references. For example ``\textit{... the end of next year, ...}" with temporal tag $1992\_06\_02$ is replaced by ``\textit{... the end of 1993, ...}".
\newline

\item \textbf{Discursive form rewriting}: ambiguous discursive forms are deleted. For example ``\textit{But, it is ...}" is replaced by ``\textit{It is ...}".
\newline

\item Finally, say clauses\footnote{For example, ambiguous say clause ``\textit{..., he said}" is removed.} and parenthesized content are removed and punctuation cleaned.

\end{enumerate}

Sentences are ordered within the summary by original document order and temporal order of documents.
Since the acronym rewriting process is dependent to the sentence order and modifies sentence's lengths, multiple passes are required to generate the final summary.
Within summary redundancy is managed by using a simple similarity threshold that prevents duplicate and highly redundant sentences to enter the summary.

\vspace*{0.3cm}

\begin{table}[h!]
\centering
\begin{tabular}{|p{.008\textwidth}|l|}
\hline

\hspace*{-0.5em} \rotatebox{90}{\hspace*{-2em} Original} &
\begin{minipage}{0.93\textwidth}
\vspace*{0.2em}
\small{\textbf{Last month}$^{\textbf{(1)}}$, U.S. scientists issued a report saying the rate of ice melting in the Arctic is increasing and within a century could lead to summertime ice-free ocean conditions not seen in the area in a million years.$^{\textbf{(2)}}$ The rate of ice melting in the Arctic is increasing and a panel of researchers says it sees no natural process that is likely to change that trend. \textbf{For example,}$^{\textbf{(3)}}$ the white sea ice reflects solar radiation back into space, but as the ice melts the dark water will absorb some of the light, warming and melting more ice.} (97 words)
\vspace*{0.3em}
\end{minipage} \\
\hline

\hspace*{-0.5em} \rotatebox{90}{\hspace*{-2.2em} Processed} &

\begin{minipage}{0.93\textwidth}
\vspace*{0.2em}
\small{The rate of ice melting in the Arctic is increasing and a panel of researchers says it sees no natural process that is likely to change that trend. The white sea ice reflects solar radiation back into space, but as the ice melts the dark water will absorb some of the light, warming and melting more ice. In 08/2005, US scientists issued a report saying the rate of ice melting in the Arctic is increasing and within a century could lead to summertime ice-free ocean conditions not seen in the area in a million years.} (95 words)
\vspace*{0.2em}
\end{minipage} \\
\hline

\end{tabular}
\caption{Example of post-processing treatments applied to the summary produced from cluster D0802A-B of TAC 2008. Dates are standardized (1); sentences are ordered with temporal constraints (2); ambiguous discursive forms are deleted (3).}
\label{tab:posttraitement}
%\vspace*{-1em}
\end{table}

% Section Experiments
\section{Experiments}
\label{sec:experiments}

The method described in the previous section has been implemented and evaluated by participating to the Text Analysis Conference (TAC) 2008 update summarization track\footnote{More information about the TAC 2008 update track is available at http://www.nist.gov/tac/} conducted by the National Institute of Standards and Technology (NIST).
The following subsections present details of the different experiments.

\subsection{The TAC 2008 update track}

Piloted in Document Understanding Conference\footnote{http://duc.nist.gov/} (DUC) 2007, the update summarization task consists in producing a short (100-word) summary of a set of newswire articles, under the assumption that the user has already read a given set of earlier articles.
The purpose of each update summary is to inform the reader of new information about a particular topic.
The test data-set in TAC 2008 comprises 48 topics.
Each topic has a topic statement (examples are given in table \ref{tab:exemple_topic}) and 20 relevant documents which have been divided into two sets\footnote{DUC 2007 data was consisting of three temporal document sets A, B and C.}: document set A and document set B.
Each document set has 10 documents, where all the documents in set A chronologically precede any of the documents in set B.
The documents are coming from the AQUAINT-2 collection of news articles.

\vspace*{0.3cm}

\begin{table}[h!]
\centering
\begin{tabular*}{0.98\textwidth}{l}
\hline
 	\textbf{Arctic and Antarctic ice melt} {\small (D0802A)} \\
\hline
	\begin{minipage}{0.95\textwidth}
	\vspace*{0.1cm}
		\textit{Describe the developments and impact of the continuing Arctic and Antarctic ice melts.}
	\vspace*{0.1cm}
	\end{minipage} \\
\hline  \\ 
 
\hline
	\textbf{Paris Riots} {\small (D0819D)} \\
\hline
	\begin{minipage}{0.95\textwidth}
	\vspace*{0.1cm}
		\textit{Describe the violent riots occurring in the Paris suburbs beginning October 27, 2005. Include details of the causes and casualties of the riots and government and police responses.}
	\vspace*{0.1cm}
	\end{minipage} \\
\hline
\end{tabular*}
\caption{Example of topic statements (D0802A and D0819D).}
\label{tab:exemple_topic}
\end{table}

%D0819D
%Paris Riots
%Describe the violent riots occurring in the Paris suburbs beginning October 27, 2005. Include details of the causes and casualties of the riots and government and police responses.

Given a DUC topic and its two document sets (A and B), the task is to create two brief, fluent summaries that contribute to satisfying the information need expressed in the topic statement.
The first one is a topic-oriented summary of the document set A while the second one is an update summary of the document set B produced under the assumption that the reader has already read documents in set A.

\subsection{Evaluation}

All summaries produced by our approach were evaluated both automatically and manually by the NIST.
The manual evaluation comprised three scores:

\begin{itemize}
\item an \emph{Overall Responsiveness} score\footnote{Integer between 1 (very poor) and 5 (very good).} based on both the linguistic quality of the summary and the amount of information in the summary that helps to satisfy the information need expressed in the topic narrative.
\item a \emph{Linguistic Quality} score$^\text{\scriptsize{3}}$ guided by consideration of the following factors: grammaticality, non-redundancy, referential clarity, focus, structure and coherence.
\item a \emph{Pyramid} \cite{nenkova2004} recall score computed on Summary Content Units (SCUs) annotations. Human annotators select overlapping content in multiple model summaries to construct a pyramid of SCUs.
\end{itemize}

Most existing automated evaluation methods work by comparing the generated summaries to one or more reference summaries (ideally, produced by humans).
In the TAC 2008 evaluation, four human summaries were written for each document set.
To evaluate the quality of our generated summaries, several automatic measures were computed:

\begin{itemize}
	\item \textsc{Rouge}\footnote{\textsc{Rouge} is available at http://haydn.isi.edu/ROUGE/} \cite{lin2004} is a $n$-gram recall measure calculated between a candidate summary and a set of reference summaries. It is computed as
			
	\begin{equation}
	\text{\textsc{Rouge}-(N)} =  
	\frac{\sum_{ s \in R_{ref} } \sum_{ \textit{N-grams} \in s} \textit{Co-occurrences}(\textit{N-grams})}
	{\sum_{s \in R_{ref}} \sum_{ \textit{N-grams} \in s} Count(\textit{N-grams})}
	\label{rouge-n}
	\end{equation}
	
	where $N$ stands for the length of the $n$-gram and \textit{Co-occurrences}(\textit{N-grams}) is the maximum number of $n$-grams co-occurring in a candidate summary and a set of reference summaries. In our experiments \textsc{Rouge-1}, \textsc{Rouge-2} and \textsc{Rouge-su4} will be computed. 
	
	\item \emph{Basic Elements}\footnote{\emph{Basic Elements} is available at http://haydn.isi.edu/BE/} \cite{hovy2006} is similar to \textsc{Rouge} but uses minimal-length fragments of \emph{sensible meaning} as units such as ``\textit{kitchen knife}" or ``\textit{Bank of America}".
\end{itemize}

In the TAC 2008, NIST received 71 runs from 33 participants for the update summarization task.
Each participant submitted up to three runs, ranked by priority.
All runs were evaluated automatically (71 runs) but manual evaluations were provided only for runs with
priority 1 and 2 (57 runs).
In addition, one baseline summarizer was included in the evaluation.
It consists in returning all the leading sentences (up to 100 words) in the most recent document.
The DUC 2007 update data was used to train our system and to estimate the interpolation coefficient of the similarity measure and the novelty factor.
As the DUC 2007 update task was consisting of three temporal documents sets, we have adapted the data set to match the TAC 2008 guideline by removing the third cluster.
Parameters for the relevance function and the novelty factor were tuned using this modified data set.
The optimal values we have found are $\alpha = 0.7$ and $\mathcal{N}\hspace*{-0.08cm}f(H) = 1/c$ with $c = 1$ for cluster A (no history) and $c = 2$ for cluster B. 
%The optimal parameters are $\alpha = 0.7$ and $f(H) = 0.5$.

\subsection{Official results}

Table \ref{tab:resultats_fusion} shows the results obtained by our submission at the update summarization task of TAC 2008.
Our system has achieved good results for Overall Responsiveness and Linguistic Quality, respectively ranked 22$^{th}$ and 14$^{th}$ out of 58 submissions, but average ones for automatic evaluations, ranked between the 42$^{th}$ and 32$^{th}$ place out of 72 submissions.
Giving more confidence to manual evaluation, we can say that our system performed quite well.
%Our performance is however handicapped by lows ranks 
One surprising result is that our system has obtained high marks in linguistic quality despite the simplicity of our rule based post-processing.

\vspace*{0.3cm}
\begin{table}[H]
\centering
\vspace{+5pt}
\begin{tabular}{lrr}
	\hline
	\textbf{Evaluation} & \textbf{Score} & \textbf{Rank}\\
	\hline
		Overall Responsiveness & 2.33 & 22/58\\
		Linguistic Quality & 2.65 & 14/58\\
		\textit{Pyramid} & 0.238 & 30/58\\
	\hline
		\textsc{Rouge-1} & 0.33611 & 42/72\\
		\textsc{Rouge-2} & 0.07450 & 38/72\\
		\textsc{Rouge-su4} & 0.11581 & 32/72\\
		\textit{Basic Elements} & 0.04574 & 35/72\\
	\hline
\end{tabular}
\vspace{+5pt}
\caption{Results of manual and automatic evaluations at the TAC 2008 update task.}
\label{tab:resultats_fusion}
\vspace{-3pt}
\end{table}

\begin{figure}[H]
	\centering
	\includegraphics[width=0.75\textwidth]{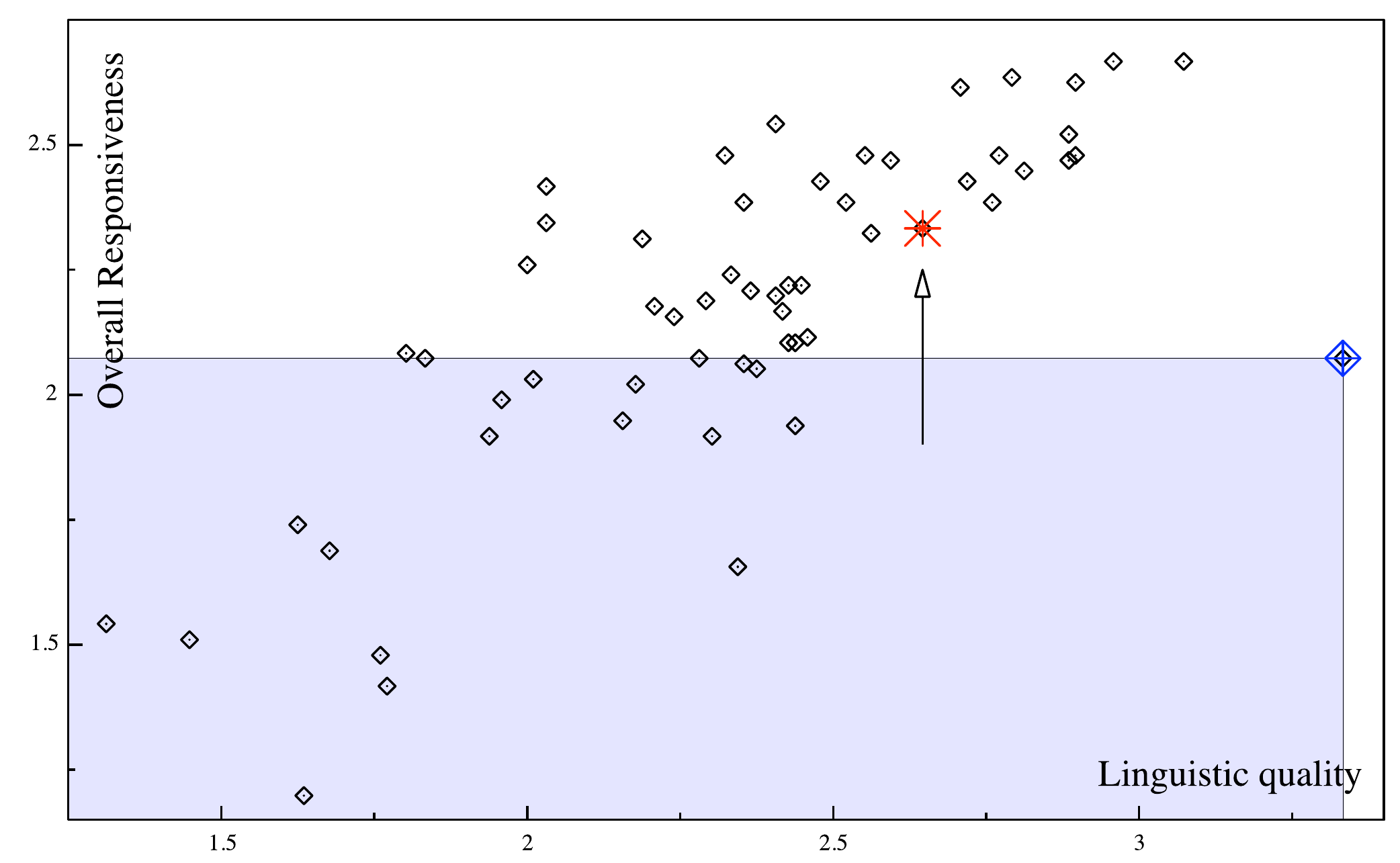}
	%\vspace{-5pt}
	\caption{Scatter plot of Linguistic quality and Overall responsiveness for the TAC 2008 update task. Our system (red star) and the baseline (big blue diamond) are highlighted.}
	\label{fig:system_human}
\end{figure}

\begin{figure}[H]
	\centering
	\includegraphics[width=0.75\textwidth]{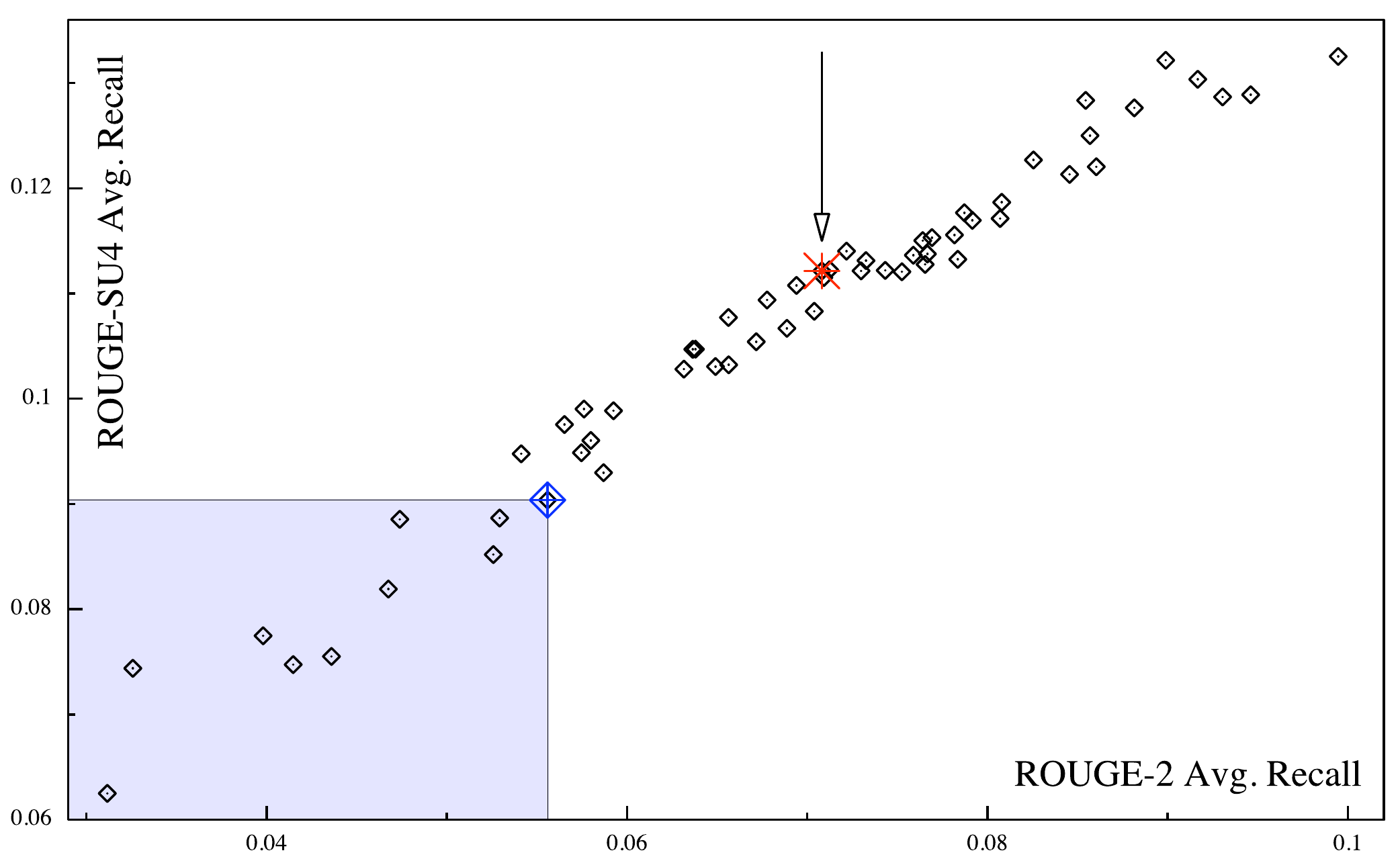}
	%\vspace{-5pt}
	\caption{Scatter plot of \textsc{Rouge-2} and \textsc{Rouge-su4} average recall scores for the TAC 2008 update task. Our system (red star) and the baseline (big blue diamond) are highlighted.}
	\label{fig:system_auto}
\end{figure}

For a comparative evaluation, Figures \ref{fig:system_human} and \ref{fig:system_auto} show the results obtained by all the systems participating in the update summarization task at TAC 2008.
The baseline consisting of 100-word summaries generated by taking the first sentences in most recent articles is also shown in the two figures.
It is worth noting that teams were allowed to submit up to three runs, generally consisting of different parameter configurations.
That way, the number of submissions that have obtained better marks than our system may have in fact been produced by a number of systems three times lower.
Being more balanced between content and linguistic evaluations, our system always outperforms the widely used lead-based baseline that have been proved to be very challenging \cite{brandow1995}.

%\vspace*{-1cm}

Results for separated document sets are presented in Table \ref{tab:difference_A_B}.
One can say that evaluation scores are significantly lower for summaries of document sets B but it is worth noting that manual evaluation ranks are significantly better (Overall Responsiveness going from 26$^{th}$ to 16$^{th}$ and Linguistic Quality from 22$^{th}$ to 9$^{th}$).
This shows that, from the linguistic quality point of view, our system is less affected by the increasing difficulty of update summarization than other approaches.

\vspace*{0.3cm}
\begin{table}[H]
\centering
\begin{tabular}{lrrrr}
	\hline
	\textbf{Evaluation} & \multicolumn{2}{r}{\textbf{Docset A}} & \multicolumn{2}{r}{\textbf{Docset B}} \vspace*{-0.15cm }\\
	 & \small{score} & \small{rank} & \small{score} & \small{rank} \vspace*{-0.05cm }\\
	\hline
		Overall Resp. & 2.417 & 26/58 & 2.250 & \bleu{16/58} \\
		Linguistic Quality & 2.458 & 22/58 & 2.833 & \bleu{9/58} \\
		\textit{Pyramid} & 0.260 & 34/58 & 0.215 & \bleu{30/58}\\
	\hline
		\textsc{Rouge-2} & \hspace{-0.2cm} 0.08125 & \bleu{36/72} & 0.06783 & 43/72 \\
		\textsc{Rouge-su4} & \hspace{-0.2cm} 0.11962 & \bleu{31/72}  & 0.11211 & 32/72 \\
	\hline
\end{tabular}
\caption{Automatic and manual evaluation results for document set A and B.}
\label{tab:difference_A_B}
\end{table}

% Sous-section des resultats aditionnels
\subsection{Additional results}
In these additional experiments, \textsc{Rouge} scores have been computed using the configuration described in the official guidelines of TAC 2008\footnote{Evaluation guidelines are available at http://www.nist.gov/tac/tracks/2008/summarization/.}.
To observe the behavior of our method on presence of noisy data, we have added in each cluster a number of random documents taken from different clusters.
Since each cluster contains 10 relevant documents, this means a 2/12 (17\%), 4/14 (29\%) and 10/20 (50\%) noise on the data sets.
Results on noisy data are given in Table \ref{tab:noise_impact}.
There is no significant performance loss on our method proving that information retrieval approaches are robust for query-oriented summarization.

\vspace*{0.3cm}
\begin{table}[H]
\centering
\begin{tabular}{lrrrr}
	\hline
	\textbf{Evaluation} & \textbf{0\%} & \textbf{17\%} & \textbf{29\%} & \textbf{50\%}\\
	\hline
		\textsc{Rouge-1} & 0.33611  & 0.33604 & 0.33585 & 0.33573\\
		\textsc{Rouge-2} & 0.07450  & 0.07450 & 0.07450 & 0.07440\\
		\textsc{Rouge-su4} & 0.11581  & 0.11579 & 0.11576 & 0.11569\\
	\hline
\end{tabular}
\caption{Comparison of \textsc{Rouge} average recall scores for our system on 17\%, 29\% and 50\% noisy TAC 2008 data.}
\label{tab:noise_impact}
\end{table}

\begin{figure}[H]
	\centering
	\includegraphics[width=0.75\textwidth]{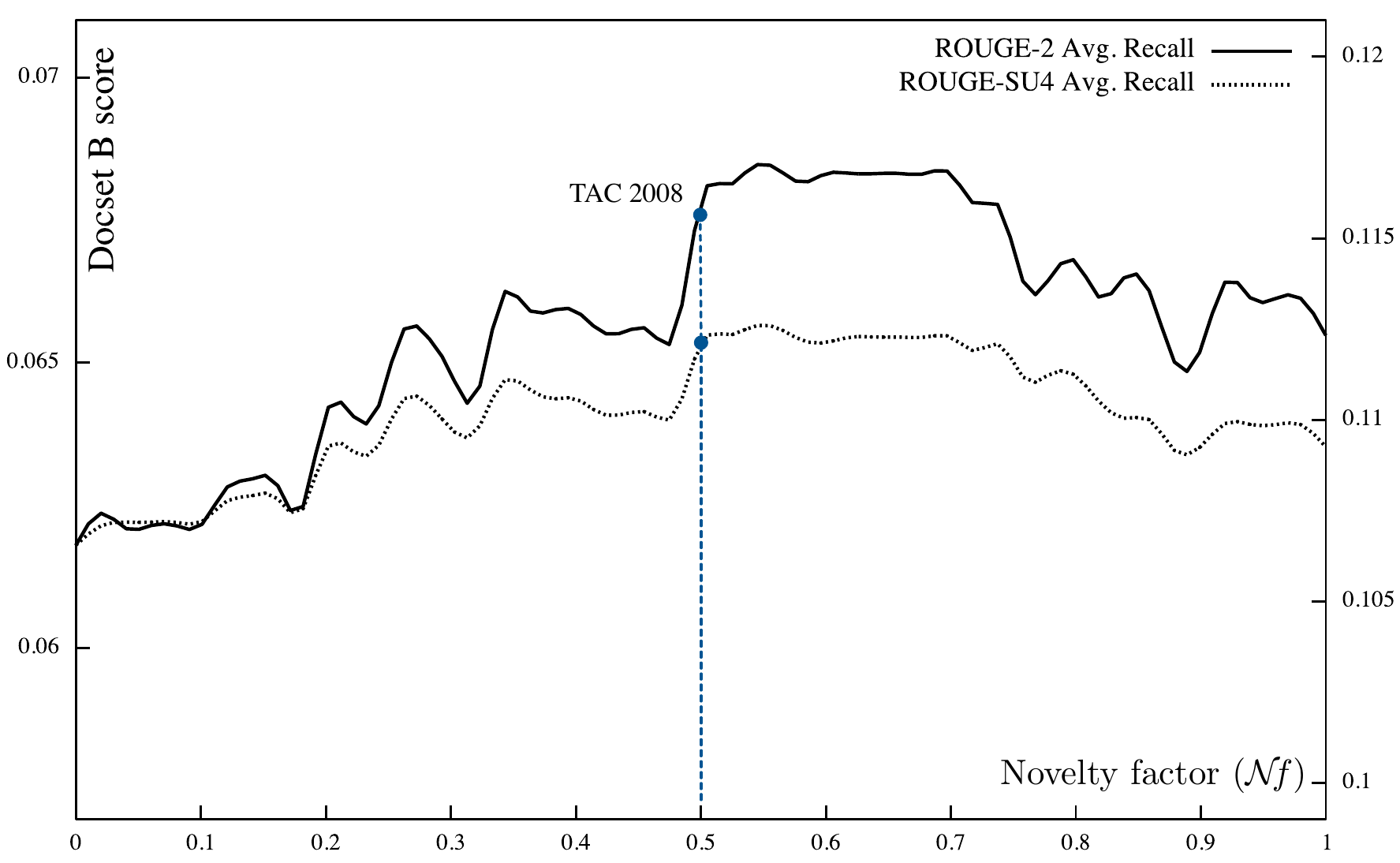}
	%\vspace{-5pt}
	\caption{Plot of \textsc{Rouge} average recall scores for docset B summaries in relation to the novelty factor $\mathcal{N}\hspace*{-0.08cm}f$ for the TAC 2008 update task.}
	\label{fig:novelty_factor}
\end{figure}

We also wanted to examine the impact of the novelty factor $\mathcal{N}\hspace*{-0.08cm}f$ used in equation (\ref{eq:smmr}) on the summaries produced for document sets B.
On Figure \ref{fig:novelty_factor}, we observe an improvement of the \textsc{Rouge} scores for all the values greater than zero, obtaining the best results for values comprised between $0.52$ and $0.68$.
The difference with the optimal value found on the training data is minimal but handicap our performance.
The size of the adapted DUC 2007 training data was obviously too small (10 topics of 18 documents) to avoid over-fitting problems.

% Section Discussion
\section{Discussion}
\label{sec:discussion}

The summarizer based on the \textsc{Smmr} sentence scoring algorithm succeeds in identifying most relevant --but containing new facts-- sentences from clusters of news articles.
The results obtained during the TAC 2008 evaluation prove that our method can achieve good results for both linguistic and content quality.
Unlike other approaches, our system does not use large linguistic or knowledge resources which makes it lightweight and easily adaptable to any other language or any domain.
Computing the whole TAC 2008 test data takes less than five minutes on a 2Ghz dual-core with 1Gb of RAM.
As applications that are subject to use update summarization algorithms are gathering tremendous amount of data such as news aggregators, computational complexity is becoming an important feature to take into consideration. 

We have observed another interesting result on our submission: automatic and manual evaluations are not often correlated.
To illustrate this lack of correlation, the topics that, within our submission, have received the best manual (D0828) and automatic (D0845) scores are compared.
Results are shown in Table \ref{tab:difference_best_worst}.
As we can see, manual and automatic evaluation scores are in total contradiction.
Indeed, according to manual evaluations, our best summaries have been generated for the topic D0828 while automatic scores for this topic are poor.
Inversely, according to automatic scores, our best topic is D0845 while its manual scores are very poor. 
By scrutinizing the generated summaries shown in the Table \ref{tab:exemple_resume_tac2008}, we have identified the reasons of this issue.
Redundancy is the main factor for these high \textsc{Rouge} scores.
Units of meaning such as ``\textit{the ivory-billed woodpecker}" are split in an incorrect way, wrongly increasing the number of matching tokens used for computing recall scores.
This example proves that using only automatic evaluations is somehow risky.

\begin{table}[H]
\centering
\begin{tabular}{lrr}
	\hline
	\textbf{Evaluation} & \textbf{D0828} & \textbf{D0845}\\
	\hline
		Overall Responsiveness & \textbf{4.0} \small{(\bleu{1})} & 1.5 \small{(35)} \\
		Linguistic Quality & \textbf{3.5} \small{(\bleu{6})} & 2.0 \small{(29)} \\
		\textit{Pyramid} & \textbf{0.324} \small{(\bleu{1})} & 0.215 \small{(38)}\\
	\hline
		\textsc{Rouge-1} & \hspace{-0.2cm}0.32993 \small{(26)} & \textbf{0.39986} \small{(\bleu{4})}\\
		\textsc{Rouge-2} & \hspace{-0.2cm}0.06995 \small{(24)} & \textbf{0.14724} \small{(\bleu{1})} \\
		\textsc{Rouge-su4} & \hspace{-0.2cm}0.11299 \small{(28)} & \textbf{0.18378} \small{(\bleu{1})}\\
		\textit{Basic Elements} & \hspace{-0.2cm}0.05562 \small{(18)} & \textbf{0.05641} \small{(\bleu{16})}\\
	\hline
\end{tabular}
%\vspace*{-0.1cm}
\caption{Results of manual and automatic evaluations for topics D0828 et D0845. Ranks obtained by the topic within our submission are shown in parenthesis. The topic ranked in first place contains the summaries that have obtained the best scores in comparison to the other topics of our submission.}
\label{tab:difference_best_worst}
\end{table}

\begin{table}[h!]
\centering

%%%%%%%%%%% EXAMPLE D0828 %%%%%%%%%%%%%%%%
\begin{tabular}{|p{.006\textwidth}|l|}

	\hline
	\multicolumn{2}{|l|}{
	\begin{minipage}{0.95\textwidth}
		\vspace*{0.2em}
		\textbf{Martha Stewart in Prison\\}
		\textit{Describe Martha Stewart's experiences while in prison.}
		\vspace*{0.2em}
	\end{minipage}
	}\\
	\hline

\hspace*{-0.5em} \rotatebox{90}{\hspace*{-2em} \textbf{D0828-A}} &

\begin{minipage}{0.95\textwidth}
\vspace*{0.2em}
\small{NEW YORK It's check-in day for Martha Stewart. Larry Stewart, who is not related to Martha Stewart, was acquitted of the charges. Q. What will happen to the company, Martha Stewart Living Omnimedia? Stewart spends up to three hours a night writing on a prison typewriter with ribbons purchased at a prison store. Bacanovic and Stewart were both given the option of staying out of prison while they appealed. Martha Stewart has been exercising, reading and making friends in prison, but the food at the minimum-security prison camp in West Virginia is "terrible," the domestic diva's daughter said.
}
\vspace*{0.2em}
\end{minipage} \\

\hline

\hspace*{-0.5em} \rotatebox{90}{\hspace*{-2em} \textbf{D0828-B}} &
\begin{minipage}{0.95\textwidth}
\vspace*{0.2em}
\small{Martha Stewart, in a Christmas message posted on her personal Web site, called for sentencing reform and took a swipe at the "bad food" in prison. Since entering federal prison in october, Martha Stewart has tried her hand at ceramics, learned to crochet and become an expert on vending-machine snacks. Martha Stewart, who is about to get out of prison, seems to have undergone a makeover on the cover of the latest Newsweek. One of the tasks ahead of Stewart is to try and spin the goodwill she gained in prison into profits for her Martha Stewart Living Omnimedia Inc.
}
\vspace*{0.2em}
\end{minipage} \\

\hline
\end{tabular}

\vspace*{0.4cm}

%%%%%%%%%%% EXAMPLE D0845 %%%%%%%%%%%%%%%%
\begin{tabular}{|p{.006\textwidth}|l|}
	
	\hline
	\multicolumn{2}{|l|}{
	\begin{minipage}{0.95\textwidth}
		\vspace*{0.2em}
		\textbf{Ivory-billed woodpecker\\}
		\textit{Describe developments in the rediscovery of the ivory-billed woodpecker, long thought to be extinct.}
		\vspace*{0.2em}
	\end{minipage}
	}\\
	\hline

\hspace*{-0.5em} \rotatebox{90}{\hspace*{-2em} \textbf{D0845-A}} &

\begin{minipage}{0.95\textwidth}
\vspace*{0.2em}
\small{The ivory-billed woodpecker, a bird long thought extinct, has been sighted in the swamp forests of eastern Arkansas for the first time in more than 60 years, Cornell University scientists said. "The ivory-billed woodpecker, long suspected to be extinct, has been rediscovered in the 'Big Woods' region of eastern Arkansas", researchers reported in the journal Science to be published. The ivory-billed woodpecker is one of six North American bird species thought to have gone extinct since 1880. The ivory-billed woodpecker, once prized for its plumage and sought by American Indians as magical, was thought to be extinct for years.
}
\vspace*{0.2em}
\end{minipage} \\

\hline

\hspace*{-0.5em} \rotatebox{90}{\hspace*{-2em} \textbf{D0845-B}} &
\begin{minipage}{0.95\textwidth}
\vspace*{0.2em}
\small{Recordings of the ivory-billed woodpecker's distinctive double-rap sounds have convinced doubting researchers that the large bird once thought extinct is still living in an east Arkansas swamp. The recordings seem to indicate that there is more than one ivory-billed woodpecker in the area. For half a century, bird-watchers have longed for a glimpse of the ivory-billed woodpecker, a bird long given up for extinct but recently rediscovered in Arkansas. The ivory-billed woodpecker was thought to be extinct until it was spotted in the swamps of southeast Arkansas in 2004. The ivory bill was, or is, the largest North American woodpecker.
}
\vspace*{0.2em}
\end{minipage} \\

\hline
\end{tabular}
%\vspace*{-0.2cm}
\caption{Examples of our submission for the topics D0828 and D0845 of TAC 2008.}
\label{tab:exemple_resume_tac2008}
\end{table}

% Section conclusions
\section{Conclusions}
\label{sec:conclusions}
%In this paper we have described an update summarization system based on \textsc{Smmr}, a scalable sentence scoring method inspired by the MMR algorithm.
In this paper we have explained how we had revisited the classical MMR algorithm in order to propose a novel approach to update summarization so called the \textsc{Smmr}.
An important aspect of our approach is that it does not requires re-ranking nor linguistic knowledge\footnote{Our system only uses minimal linguistic resources for post-processing that are easily adaptable to any other language.}, which makes it a simple and efficient method to tackle the issue of update summarization.

The novelty factor, characterized in our sentence scoring method by a linear function $\mathcal{N}\hspace*{-0.08cm}f(H)$, turns out to be a very important parameter requiring to be tuned in a more judicious manner.
Using a linear function that relies on the number of previous clusters instead of the exact amount of text can be hazardous.
High redundancy within news articles forces us to believe that the reader can gain knowledge of only a reduced number of concepts. 
This is the reason why we think computing the novelty factor by using the concept redundancy is worthy of further work.
Recent work by \cite{thadani2008coling} gives some interesting ideas on how to remove redundancy by constructing novel graph-based representations from documents.

It was pointed out that Question Answering and query-oriented summarization have been converging on a common task, the value added by summarization lying in the linguistic quality.
We have seen that applying simple ruled-based linguistic treatments to candidate sentences allows to significantly increase the linguistic quality.

Current research works are predominantly focused on the English language.
This is why we are currently developing a bilingual evaluation corpus (English and French).
Among the others, this point sounds like a promise for further investigation.

\section*{Acknowledgments}
This work was supported by the \textit{Agence Nationale de la Recherche}, France, project RPM2. http://labs.sinequa.com/rpm2/

\bibliography{boudin_smmr09}

\begin{thebibliography}{10}

\bibitem{allan2001}
J.~Allan, R.~Gupta, and V.~Khandelwal.
\newblock {Temporal summaries of new topics}.
\newblock In {\em Proceedings of the 24th annual international ACM SIGIR
  conference on research and development in information retrieval}, pages
  10--18. ACM New York, NY, USA, 2001.

\bibitem{boudin_thesis}
F.~Boudin.
\newblock {\em {Exploration d'approches statistiques pour le r\'esum\'e
  automatique de texte}}.
\newblock PhD thesis, Universit\'e d'Avignon et des pays de Vaucluse, December
  2008.

\bibitem{boudin_coling08}
Florian Boudin, Marc El-B\`{e}ze, and Juan-Manuel Torres-Moreno.
\newblock A scalable {MMR} approach to sentence scoring for multi-document
  update summarization.
\newblock In {\em Coling 2008: Companion volume: Posters and Demonstrations},
  pages 21--24, Manchester, UK, August 2008. Coling 2008 Organizing Committee.

\bibitem{boudin_ranlp07}
Florian Boudin and Juan-Manuel Torres-Moreno.
\newblock {A Cosine Maximization-Minimization approach for User-Oriented
  Multi-Document Update Summarization}.
\newblock In {\em Recent Advances in Natural Language Processing (RANLP)},
  pages 81--87, Borovets, Bulgaria, September 2007.

\bibitem{boudin_gotal08}
Florian Boudin, Juan-Manuel Torres-Moreno, and Patricia Vel\'{a}zquez-Morales.
\newblock {An Efficient Statistical Approach for Automatic Organic Chemistry
  Summarization}.
\newblock In Bengt Nordstr\"{o}m and Aarne Ranta, editors, {\em 6th
  International Conference on Natural Language Processing, Go{TAL} 2008},
  volume 5221 of {\em Lecture Notes in Computer Science}, pages 89--99,
  Gothenburg, Sweden, August 2008. Springer.

\bibitem{brandow1995}
R.~Brandow, K.~Mitze, and L.F. Rau.
\newblock {Automatic condensation of electronic publications by sentence
  selection}.
\newblock {\em Information Processing and Management}, 31(5):675--685, 1995.

\bibitem{carbonell1998}
J.~Carbonell and J.~Goldstein.
\newblock {The use of MMR, diversity-based reranking for reordering documents
  and producing summaries}.
\newblock In {\em 21st annual international ACM SIGIR conference on Research
  and development in information retrieval}, pages 335--336. ACM Press New
  York, NY, USA, 1998.

\bibitem{daume_thesis}
H.~Daum{\'e}~III.
\newblock {\em {Practical Structured Learning for Natural Language
  Processing}}.
\newblock PhD thesis, University of Southern California, August 2006.

\bibitem{edmundson1969}
H.~P. Edmundson.
\newblock {New methods in automatic extracting}.
\newblock {\em Journal of the ACM (JACM)}, 16(2):285, 1969.

\bibitem{erkan2004}
G.~Erkan and D.R. Radev.
\newblock {LexRank: Graph-based Lexical Centrality as Salience in Text
  Summarization}.
\newblock {\em Journal of Artificial Intelligence Research}, 22(2004):457--479,
  2004.

\bibitem{hachey2005}
B.~Hachey, G.~Murray, and D.~Reitter.
\newblock {The Embra System at DUC 2005: Query-oriented Multi-document
  Summarization with a Very Large Latent Semantic Space}.
\newblock In {\em Document Understanding Conference (DUC)}, Vancouver, Canada,
  October 2005.

\bibitem{hickl_duc07}
A.~Hickl, K.~Roberts, and F.~Lacatusu.
\newblock {LCC's GISTexter at DUC 2007: Machine Reading for Update
  Summarization}.
\newblock In {\em Document Understanding Conference (DUC)}, Rochester, USA,
  April 2007.

\bibitem{Hinrichs1986}
Erhard Hinrichs.
\newblock Temporal anaphora in discourses of english.
\newblock {\em Linguistics and Philosophy}, 9(1):63--82, February 1986.

\bibitem{hovy2006}
E.~Hovy, C.Y. Lin, L.~Zhou, and J.~Fukumoto.
\newblock {Automated Summarization Evaluation with Basic Elements}.
\newblock In {\em Fifth Conference on Language Resources and Evaluation
  (LREC)}, May 2006.

\bibitem{kupiec1995}
J.~Kupiec, J.~Pedersen, and F.~Chen.
\newblock {A trainable document summarizer}.
\newblock In {\em Proceedings of the 18th annual international ACM SIGIR
  conference on Research and development in information retrieval}, pages
  68--73. ACM New York, NY, USA, 1995.

\bibitem{ledeneva_micai08}
Yulia Ledeneva.
\newblock Effect of preprocessing on extractive summarization with maximal
  frequent sequences.
\newblock In {\em {MICAI} 2008: Advances in Artificial Intelligence}, volume
  5317/2008 of {\em Lecture Notes in Computer Science}, pages 123--132.
  Springer Berlin / Heidelberg, 2008.

\bibitem{li_coling08}
Wenjie Li, Furu Wei, Qin Lu, and Yanxiang He.
\newblock {PNR2}: Ranking sentences with positive and negative reinforcement
  for query-oriented update summarization.
\newblock In {\em Proceedings of the 22nd International Conference on
  Computational Linguistics (Coling 2008)}, pages 489--496, Manchester, UK,
  August 2008. Coling 2008 Organizing Committee.

\bibitem{lin2004}
Chin-Yew Lin.
\newblock Rouge: A package for automatic evaluation of summaries.
\newblock In Stan~Szpakowicz Marie-Francine~Moens, editor, {\em Text
  Summarization Branches Out: Proceedings of the ACL-04 Workshop}, pages
  74--81, Barcelona, Spain, July 2004. Association for Computational
  Linguistics.

\bibitem{lin_duc07}
Z.~Lin, T.S. Chua, M.Y. Kan, W.S. Lee, L.~Qiu, and S.~Ye.
\newblock {NUS at DUC 2007: Using Evolutionary Models of Text}.
\newblock In {\em Document Understanding Conference (DUC)}, Rochester, USA,
  April 2007.

\bibitem{luhn1958}
H.~P. Luhn.
\newblock {The automatic creation of literature abstracts}.
\newblock {\em IBM Journal of Research and Development}, 2(2):159--165, 1958.

\bibitem{mani2002}
I.~Mani, G.~Klein, D.~House, L.~Hirschman, T.~Firmin, and B.~Sundheim.
\newblock {SUMMAC: a text summarization evaluation}.
\newblock {\em Natural Language Engineering}, 8(01):43--68, 2002.

\bibitem{mani1999}
I.~Mani and M.T. Maybury.
\newblock {\em {Advances in Automatic Text Summarization}}.
\newblock MIT Press, 1999.

\bibitem{murray2005}
G.~Murray, S.~Renals, and J.~Carletta.
\newblock {Extractive Summarization of Meeting Recordings}.
\newblock In {\em Ninth European Conference on Speech Communication and
  Technology (Eurospeech)}, Lisboa, Portugal, September 2005.

\bibitem{nenkova2004}
Ani Nenkova and Rebecca Passonneau.
\newblock Evaluating content selection in summarization: The pyramid method.
\newblock In Daniel~Marcu Susan~Dumais and Salim Roukos, editors, {\em
  HLT-NAACL 2004: Main Proceedings}, pages 145--152, Boston, Massachusetts,
  USA, May 2004. Association for Computational Linguistics.

\bibitem{radev1998}
D.R. Radev and K.R. McKeown.
\newblock {Generating natural language summaries from multiple on-line
  sources}.
\newblock {\em Computational Linguistics}, 24(3):469--500, 1998.

\bibitem{salton1975}
G.~Salton, A.~Wong, and C.~S. Yang.
\newblock {A vector space model for automatic indexing}.
\newblock {\em Communications of the ACM}, 18(11):613--620, 1975.

\bibitem{jones1972}
K.~Sp\"arck~Jones.
\newblock {A statistical interpretation of term specificity and its application
  in retrieval}.
\newblock {\em Journal of Documentation}, 28(1):11--21, 1972.

\bibitem{teufel1997}
S.~Teufel and M.~Moens.
\newblock {Sentence extraction as a classification task}.
\newblock In {\em ACL/EACL workshop on” Intelligent and scalable Text
  summarization}, pages 58--65, 1997.

\bibitem{thadani2008coling}
Kapil Thadani and Kathleen McKeown.
\newblock A framework for decreasing textual redundancy.
\newblock In {\em Coling 2008}, Manchester, UK, August 2008. Coling 2008
  Organizing Committee.

\bibitem{winkler1999}
W.E. Winkler.
\newblock {The state of record linkage and current research problems}.
\newblock {\em Statistics of Income Division, Internal Revenue Service
  Publication R}, 4, 1999.

\bibitem{witte_duc07}
Ren{\'{e}} Witte, Ralf Krestel, and Sabine Bergler.
\newblock {Generating Update Summaries for DUC 2007}.
\newblock In {\em Document Understanding Conference (DUC)}, Rochester, USA,
  April 2007.

\bibitem{ye2005}
S.~Ye, L.~Qiu, T.S. Chua, and M.Y. Kan.
\newblock {NUS at DUC 2005: Understanding documents via concept links}.
\newblock In {\em Document Understanding Conference (DUC)}, Vancouver, Canada,
  October 2005.

\end{thebibliography}
\bibliographystyle{plain}

\end{document}